%% file: correlated-random-access-schemes.tex
\documentclass[conference,10pt]{IEEEtran}
\usepackage{amsmath}
\usepackage{amssymb}
\usepackage{bbm}
\usepackage[capitalize]{cleveref}
\usepackage{pgfplots}
\pgfplotsset{compat=1.14}
\usetikzlibrary{patterns}
\usetikzlibrary{plotmarks}
\usepgfplotslibrary{groupplots}
\usepackage{algorithm}
\usepackage{algorithmic}
\usepackage{url}

\begin{document}
\title{Random Access Schemes in Wireless Systems With Correlated User Activity}

\author{\IEEEauthorblockN{Anders Ellersgaard Kal{\o}r, Osama A. Hanna, Petar Popovski}
\IEEEauthorblockA{Department of Electronic Systems,
Aalborg University, Denmark\\
Email: \{aek,habib,petarp\}@es.aau.dk}
}

\maketitle

\begin{abstract}
Traditional random access schemes are designed based on the aggregate process of user  activation, which is created on the basis of independent activations of the users. However, in Machine-Type Communications (MTC), some users are likely to exhibit a high degree of correlation, e.g. because they observe the same physical phenomenon. This paves the way to devise access schemes that combine scheduling and random access, which is the topic of this work. The underlying idea is to schedule highly correlated users in such a way that their transmissions are less likely to result in a collision. To this end, we propose two greedy allocation algorithms. Both attempt to maximize the throughput using only pairwise correlations, but they rely on  different assumptions about the higher-order dependencies. We show that both algorithms achieve higher throughput compared to the  traditional random access schemes, suggesting that user correlation can be utilized effectively in access protocols for MTC. 
\end{abstract}
 

\section{Introduction}
Machine-Type Communication (MTC) represents an important pillar of 5G wireless systems. It will come in two flavors, massive Machine Type Communications (mMTC) and Ultra-Reliable and Low Latency Communications (URLLC) services.
In contrast to traditional uses of communication systems dominated by bandwidth intensive, human-initiated activity, machine-type traffic is characterized by a very large number of devices, small packet sizes and possibly strict latency and reliability requirements~\cite{Boccardi}. Furthermore, machines are likely to produce more correlated and predictable traffic patterns, e.g. if the traffic is generated based on observations of some common physical phenomenon~\cite{3gpp_tr_37868}.

Current access protocols do not exploit this correlation between users, and are usually designed based on the aggregate activation process under the assumption that users are independent. Under these conditions, the access protocols are derivatives of slotted ALOHA~\cite{Roberts:1975:APS:1024916.1024920} which achieves maximum throughput per slot of $1/e\approx 0.37$ when the average number of transmissions per slot is 1. The recent class of random access protocols that rely on successive interference cancellation (SIC), such as coded slotted ALOHA~\cite{7302046,6630481}, can achieve high throughputs at the expense multiple packet replicas sent by the users and complex processing at the receiver. Regardless of the receiver model, none of these protocols considers correlation of the activity among the transmitting devices (users).

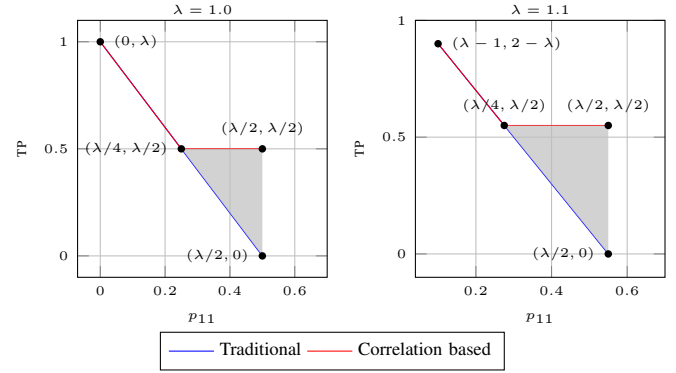
\begin{figure}
\centering
\input{motivating_example_lambda1.tex}
\input{motivating_example_lambda11.tex}
\ref{named}
\caption{Achievable throughputs with two correlated users and different arrival rates $\lambda$. The gray area represents the throughput achievable by exploiting correlation statistics.}
\label{fig:motivating_example}
\vspace{-6pt}
\end{figure}

In this paper, we investigate how knowledge about the user activity correlation can be used to improve the throughput of the access protocols, and consequently reduce the latency and increase the reliability of the systems. In this first work on the topic, we limit ourselves to the collision model, but the ideas can be extended to the receivers based on SIC. To illustrate  the main idea, consider an example of a system with $N=2$ users and $K\in\{1,2\}$ slots. Let $X_{1}$ and $X_{2}$ denote the random binary events that user 1 and user 2 transmits in a given frame, and let $p_{ij}=\Pr(X_{1}=i,X_{2}=j)$ with $i,j\in \{0,1\}$. For simplicity, assume that $p_{01}=p_{10}\triangleq p$, such that the expected number of transmissions is $\lambda = 2(p+p_{11})$.
If $K=1$ the two users contend for the slot, and a transmission will succeed only in the case where a single user transmits. On the other hand, if $K=2$ the user transmissions will succeed unconditionally.
The throughput is given by
\begin{equation*}
\mathrm{TP}=\frac{2p+2p_{11}\mathbbm{1}(K>1)}{K}
\end{equation*}
where $\mathbbm{1}(K>1)$ equals $1$ if $K>1$ and $0$ otherwise.
Under the traditional assumption of independent users $p_{11}=(\lambda/2)^2$. Since by definition $p=\lambda/2-p_{11}$ the optimal policy is to allocate two slots only when $\lambda>1$.
However, if the users are correlated, i.e. $p_{11}\neq (\lambda/2)^2$, this is a suboptimal strategy that leads to reduced utilization if $p_{11}<(\lambda/2)^2$ or increased collision probability if $p_{11} >(\lambda/2)^2$. If we instead make use of $p_{11}$, the optimal strategy is to allocate 2 slots when $\lambda/4<p_{11}$.
This is illustrated in \cref{fig:motivating_example} for $\lambda=1.0$ and $\lambda=1.1$, where the gray region indicates the throughput gain that can be achieved by considering the user correlation. Notice that the two schemes perform equivalently in the region where $p_{11}$ is small, since allocating a single slot is optimal in both schemes.

In \cref{sec:sysmodel} we generalize the allocation problem to $N$ users and $K$ slots and define the system model. Since this turns out to be non-convex and hard to solve even approximately, we reformulate the problem and propose two heuristic allocation algorithms in \cref{sec:heuristicalg}. In \cref{sec:trafficmodels} we define a traffic model that we use in \cref{sec:results} to evaluate the algorithms.
Finally, we discuss practical aspects and future work in \cref{sec:practicalaspects} and conclude the paper in \cref{sec:conclusion}.

\section{System Model and Problem Definition}\label{sec:sysmodel}

We now generalize the scenario considered in the previous section and consider $N$ users that transmit to a common receiver in frames consisting of $K$ slots. We consider a collision model, such that when there are two or more transmissions in a slot, the receiver observes a failure (erasure). In each frame, each user can have at most one transmission, as in the original framed ALOHA~\cite{Roberts:1975:APS:1024916.1024920}.

The user activation is defined through the joint distribution $\Pr(\mathbf{x})=\Pr(x_1,x_2,\ldots,x_N)$ where $x_n=1$ if user $n$ transmits in a given slot, and $x_n=0$ otherwise. It is assumed that this joint distribution is independently sampled in each new $K-$slot frame.
As seen in the example with $N=2$ users, the correlation can be used to decide which users should be assigned to the same slot. Thus, in practical scenarios where the number of slots is much smaller than the number of users, the overall objective is to assign the users to the slots so as to maximize the throughput. 
We define the allocation matrix $\mathbf{A}\in\mathbb{R}^{N\times K}$ where $A_{ij}$ is the probability that user $i$ transmits in slot $j$ conditioned on activation, and $\sum_j A_{ij}=1$. Let $e_{ij}$ denote the event that user $i$ selects slot $j$, i.e. $\Pr(e_{ij})=E[e_{ij}]=A_{ij}$.
The throughput is given by the expected fraction of slots in which exactly one user transmits:
\begin{equation}\label{eq:tp_optimal}
\mathrm{TP}(\mathbf{A})=\sum_{k=1}^K\sum_{n=1}^N T_n^{(k)}
\end{equation}
where
\begin{align*}
T_n^{(k)}&=E_{\mathbf{x}}\left[x_{n}E[e_{nk}]\prod_{m=1}^N(1-x_{m}E[e_{mk}])^{\mathbbm{1}(n\neq m)}\right]\\
&=E_{\mathbf{x}}\left[x_{n}A_{nk}\prod_{m=1}^N(1-x_{m}A_{mk})^{\mathbbm{1}(n\neq m)}\right].
\end{align*}
Here, we used the fact that $x_n$ and $e_{ij}$ are independent.
$\mathrm{TP}(\mathbf{A})$ is non-convex and finding the slot assignments $A_{ij}$ that maximizes this throughput is hard.
Furthermore, computing the expectations requires the full activity distribution or at least estimates of the product expectations of any subset of the users, which are in general unknown and need to be learned. The specification of the joint distribution $\Pr(x_1,\ldots,x_N)$ requires, in general, specification of $2^N-1$ values. Hence, it can only be specified (estimated) when $N$ is low. On the other hand, the number of pairwise correaltions scales as $N^2$ and can be considered feasible for estimation. We use the knowledge of pairwise correlations to resort to alternative formulations or heuristics that approximately maximize the throughput.

\begin{figure}
\centering
\input{venndiagram.tex}
\caption{Illustration of the events that contribute to the calculation of the expected throughput of user 1 in the case of four users. The hatched region indicates the expected throughput conditioned on a successful transmission.}
\label{fig:venndiagram}
\vspace{-6pt}
\end{figure}
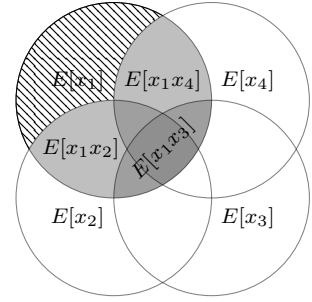

Without loss of generality, we consider the throughput contribution by user $1$ in slot $k$, $T_1^{(k)}$.
As $x_n$ are binary random variables, the expectation equals the probability of the event, i.e.
\begin{equation*}
E[x_1x_2\ldots x_N]=\Pr\left(x_1,x_2,\ldots, x_N\right).
\end{equation*}
Using this, we may express $T_1^{(k)}$ as the probability that only user $1$ transmits in slot $k$.
It follows from the inclusion-exclusion principle~\cite{flajolet2009analytic} that
the terms that include expectations of higher order than two compensate for intersecting events as illustrated graphically in \cref{fig:venndiagram} for $T_1^{(k)}$ in the case of 4 users.
Estimating the higher order expectations poses a challenge, and a reasonable objective is to only estimate the first and second order expectations, and put assumptions on the higher order expectations. Since the exact throughput for a given allocation cannot be determined without the higher order expectations, we may instead use the inclusion-exclusion principle to bound the throughput as
\begin{align}
  T_1^{(k)}&\le A_{1k}E[x_1]-\max_{m=2,\ldots,N}A_{1k}A_{mk}E[x_1x_m]\label{eq:upper}\\
  T_1^{(k)}&\ge A_{1k}E[x_1]-\sum_{m=2}^NA_{1k}A_{mk}E[x_1x_m].\label{eq:lower}
\end{align}
The lower bound is valid with equality if the higher order expectations of the users that are allocated in slot $k$ are zero, i.e. no more than two transmissions occur in the same slot. Similarly, the upper bound has equality either when the users never transmit in the same slot, or when all users $2,\ldots,N$ always transmit jointly with user 1. Both bounds may be used to derive allocations that approximate the optimal throughput. However, the lower bound provides an accurate estimation if the users are unlikely to transmit together, while the upper bound is more accurate when the users are highly correlated.

\section{Heuristic Algorithms}\label{sec:heuristicalg}
Here we present two heuristic slot assignment algorithms that attempt to maximize the throughput using only pairwise expectations based on the throughput upper/lower bounds.

\subsection{Min-Max Pairwise Correlation}
We first attempt to maximize the throughput using the upper bound (\cref{eq:upper}). Since we maximize the upper bound, we optimize for the best (in terms of probability of joint transmission) of the set of generating distributions, namely when all transmissions involving more than a one user happen in the same frame. Hence, we expect it to perform well if the users are strongly correlated in the higher order expectations, and poorly when they are anticorrelated. Considering the constraint $\sum_j A_{ij}=1$ we equivalently minimize:
\begin{equation}\label{eq:mmpc}
\mathrm{MMPC}(\mathbf{A})=\sum_{k=1}^K\max_{\{n,m\}\in [1,N]^2} A_{nk}A_{mk}E\left[x_nx_m\right]
\end{equation}
where $[1,N]^2=\{\{a,b\}:a,b\in 1,2,\ldots,N,a\neq b\}$. While \cref{eq:mmpc} simplifies \cref{eq:tp_optimal} it still constitutes a non-convex quadratically constrained linear program. While methods to approximate such problems have been proposed in the literature, including semidefinite relaxation~\cite{luo2010semidefinite,poljak1995} and successive convex approximation~\cite{beck2010,scutari2017}, they require a considerable amount of fine-tuning, or they work by lifting the variables to a higher dimension which is inappropriate for the problem at hand where the number of variables is already large.

Evaluating the performance of various optimization algorithms is beyond the scope of this paper. Instead, we focus on the case where $A_{ij}\in\{0,1\}$, i.e. users are assigned a single slot in which they deterministically transmit if they are active, and propose a simple greedy algorithm that consecutively assigns users that are less likely to transmit together jointly to the same slots. The algorithm, outlined in \cref{alg:mmpc}, takes a symmetric matrix $\mathbf{C}$ where element $ij$ is $E[x_ix_j]$ if $i\neq j$ and $\infty$ if $i=j$, and outputs the matrix $\mathbf{A}$. Here $\mathbf{C}$ can be seen as an adjacency matrix for a fully connected graph where the vertices are slots and the edge weights are the product expectation of two slots. Initially, each user is assigned to its own slot. As the algorithm proceeds, $\mathbf{C}$ is  reduced by \emph{merging} two vertices until it contains only $K$ vertices. A merge between vertex $i$ and $j$ is performed by updating the edge weights of vertex $j$: $C_{jn}=C_{nj}=\textsc{max}\{C_{ni},C_{nj}\}\,\forall n$, so that the new edge weight represents the maximum joint transmission probability between two users that are assigned to different slots. Then, the $i$-th column and row of $\mathbf{C}$ are removed (denoted by $\mathbf{C}=\mathbf{C}_{-i}$). A vector $\mathbf{S}$ that maps the users to each of the $K$ slots is maintained in order to construct the (binary) $\mathbf{A}$ matrix in the last step. In lines 9--12 the user-slot mapping is updated as a slot has been removed due to a merge. The initial size of $\mathbf{C}$ is $N\times N$, and hence the outer loop has $\mathcal{O}(N)$ iterations. The complexity within the loop is dominated by the $\textsc{min}(\mathbf{C})$ operation with $\mathcal{O}(N\log N)$, and hence the total complexity is $\mathcal{O}(N^2\log N)$.

\begin{algorithm}
\caption{MMPC allocation}
\label{alg:mmpc}
\algsetup{linenosize=\tiny}
\scriptsize
\begin{algorithmic}[1]
\renewcommand{\algorithmicrequire}{\textbf{Input:}}
\renewcommand{\algorithmicensure}{\textbf{Output:}}
\REQUIRE $\mathbf{C}\in\mathbb{R}^{N\times N},K$
\ENSURE $\mathbf{A}\in \mathbb{R}^{N\times K}$
\STATE $\mathbf{S}=[1,2,\ldots,N], \mathbf{A}=\mathbf{0}$
\WHILE {$\textsc{size}(C) > K\times K$}
\STATE $(i,j)=\textsc{min}(C)$
\FOR {$n=1,\ldots,\textsc{rows}(C)$}
\STATE $C_{jn} = \textsc{max}\{C_{ni},C_{nj}\}$
\STATE $C_{nj} = C_{jn}$
\ENDFOR
\STATE $\mathbf{C}=\mathbf{C}_{-i}$
\STATE $S_i=\textsc{min}\{S_i,S_j\}$
\FOR {$n=i,\ldots,N$}
\STATE $S_n = S_n-1$
\ENDFOR
\ENDWHILE
\FOR {$n=i,\ldots,N$}
\STATE $A_{nS_n} = 1$
\ENDFOR
\RETURN $\mathbf{S}$
\end{algorithmic}
\end{algorithm}

The active users transmit unconditionally in their assigned slots, which may result in collisions if the users are active simultaneously. To avoid this situation, we can scale the resulting allocation to maximize the probability of singular transmissions. However, as we do not know the joint activity distribution, we instead set the expected number $N_i$ of transmissions in the slot assigned to user $i$ conditioned on activity of user $i$ to one. Let $S_i$ denote the slot of user $i$ and $A_{i}$ the probability that user $i$ will transmit conditioned on being active. We then have
\begin{equation*}
  E[N_{i}]=A_{i} + \sum_{n\in\mathcal{N}_{S_i}} A_{n}E[x_n|x_i]=A_{i} + \frac{\sum_{n\in\mathcal{N}_{S_i}} A_{n}E[x_nx_i]}{E[x_i]},
\end{equation*}
where $\mathcal{N}_{S_i}$ is the set of users assigned to slot $S_i$. For each slot $j$ we determine the new values $0\le A_{i}\le 1$ that minimizes the least-squares $\sum_{i\in\mathcal{N}_j} (E[N_{i}]-1)^2$.
Although the performance the heuristic depends on the higher-order correlation of the users, results presented in \cref{sec:results} suggest that it works as intended.

\subsection{Min-Sum Pairwise Correlations}
We now consider the lower bound (\cref{eq:lower}) and minimize the sum of product expectations in each slot:
\begin{equation}\label{eq:mpc}
\mathrm{MPC}(\mathbf{A})=\frac{1}{K}\sum_{k=1}^K\sum_{n=1}^N \sum_{m=1}^N \mathbf{1}(n\neq m)A_{nk}A_{mk}E\left[x_nx_m\right].
\end{equation}
Compared to the Min-Max Pairwise Correlation, this function maximizes the throughput under the assumption that no more than two users transmit in the same slot (the higher order expectations are zero). Similar to \cref{eq:mmpc}, this also poses a non-convex quadratic program and we use a similar greedy algorithm to approximate a solution. Since we are aiming at minimizing the sum, the weight updating step in \cref{alg:mmpc} (line 5) can simply be replaced by the sum $C_{in} = C_{ij} + C_{ni} + C_{nj}$, while the remaining procedure remains unchanged.

\subsection{Illustrative Allocation Examples}
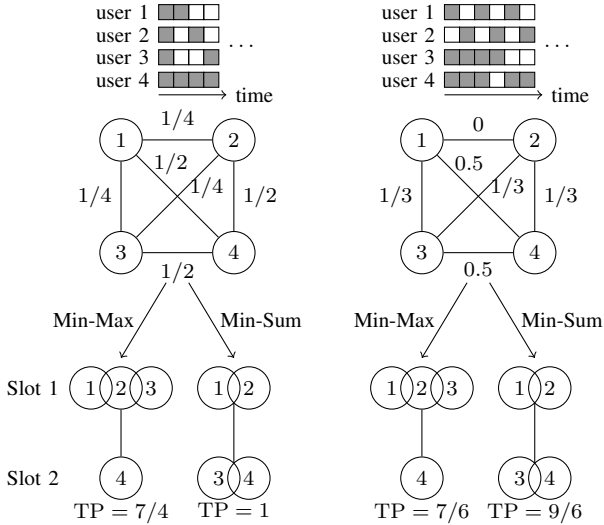
\begin{figure}
\centering
\input{illustrative_algo.tex}
\caption{Examples where the Min-Max and Min-Sum algorithms result in allocations with different throughputs. A filled square in the top indicates that the user transmits in the given frame.}
\label{fig:illustrative_algo}
\vspace{-6pt}
\end{figure}

To illustrate how the two algorithms are different, we consider two example correlations; one for which the Min-Max results in a higher throughput than the Min-Sum algorithm, and one for which the Min-Sum performs better.
We consider a scenario with four users and two slots. The first instance, for which the Min-Max algorithm is best, is illustrated on the left in \cref{fig:illustrative_algo}, where the activity pattern in the top is repeated indefinitely. The graph is defined from the input matrix $\mathbf{C}$. The Min-Max algorithm allocates users 1, 2 and 3 to slot 1, and user 4 to slot 2, resulting in a throughput of $7/4$. On the other hand, the Min-Sum algorithm allocates users 1 and 2 to slot 1, and user 3 and 4 to slot 2, yielding a throughput of only $1$. Hence, the Max-Sum algorithm achieves 75\% higher throughput. In the right graph in \cref{fig:illustrative_algo}, the opposite is the case, and the Max-Min allocation results in a throughput of $7/6$ while the Max-Sum allocation achieves $9/6$. The reason for the difference in the performance is the higher order correlations. Specifically, in the left case the higher order correlations are more significant, and hence the upper bound approximation is more accurate than the lower bound. Similarly, the lower bound approximation is more accurate in the case where the Min-Sum algorithm performs best.

\section{Evaluation}

\subsection{Traffic Model for Correlated Activity}
\label{sec:trafficmodels}
We first present the traffic model that will be used to evaluate the algorithms. The existing traffic models are not 
suitable for this study. The 3GPP model~\cite{3gpp_tr_37868} uses a $\mathrm{Beta}(3,4)$ distribution to model the arrival process at the base station. However, since this model describes the aggregate arrival process, it cannot be used for scheduling of the individual users. The activity of individual users is explicitly modeled in~\cite{laner2013traffic} through a Coupled Markov Modulated Poisson Processes, where each user is modeled as a Markov chain with transition probabilities defined as a convex combination of a user-local and a central (shared) process. While this exhibits some correlation, it acts at a macro-level and the individual users are still approximately independent at small time scales.

In the model we consider, the users are deployed uniformly in a square area in which events are generated according to a spatio-temporal Poisson point process. The users transmit then if an event occurs within a certain radius.
Let $\mathcal{R}=[0,L]^2$ denote the square region of size $L\times L$ and let $\mathbf{x}_i\sim \mathrm{Uniform}(\mathcal{R})$ denote the location of user $i$ ($i=1,\ldots,N$). We model random spatio-temporal events by a homogeneous Poisson point process on $\mathcal{R}$ with rate $\lambda$ so that the number of events within a region $\mathcal{D}\subseteq \mathcal{R}$ follows a Poisson process with rate $\lambda A(\mathcal{D})$ where $A(\mathcal{D})$ is the area of $\mathcal{D}$.

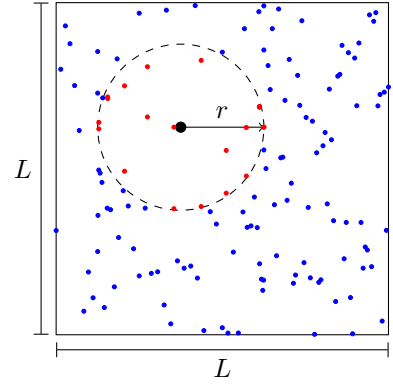
\begin{figure}
\centering
\input{spatiotemp_traffic_model.tex}
\caption{Illustration of the spatio-temporal traffic model where the black circle indicates an event, and the red dots are active devices.}
\label{fig:trafficmodel}
\vspace{-6pt}
\end{figure}

The user activity is defined in such a way that all users within radius of $r$ from an event transmit synchronously in the following frame, as illustrated in \cref{fig:trafficmodel}. The correlation between two users is defined by the intersecting area between the disks with radius $r$ centered at the users. To overcome edge effects caused by regions outside $\mathcal{R}$, we apply the border method in generation of user locations and regenerate locations that are closer than $r$ to the edge of $\mathcal{R}$. The area of the disk intersection is
$\mathcal{D}_{ij}=2r^2\cos^{-1}\left(\frac{d_{ij}}{2r}\right)-\frac{d_{ij}}{2}\sqrt{4r^2-d_{ij}^2}$, 
where $d_{ij}=\|\mathbf{x}_i-\mathbf{x}_j\|_2$, see ~\cite{weissteincirclecircle}. Two users transmit in the same frame only if an event occurs within this area, or if one or more events occur within each user's radius.
We assume that the frame duration is 1 and that a user transmits at most once per frame. It follows that $E[x_i]=1-e^{-\lambda\pi r^2}$ and for $i\neq j$
\begin{equation*}
  E[x_ix_j]=
    \begin{cases}
    \left(1-e^{-\lambda \pi r^2}\right)^2 & d_{ij}\ge r\\
    1-e^{-\lambda \mathcal{D}_{ij}} + \left(1-e^{-\lambda (\pi r^2-\mathcal{D}_{ij})}\right)^2 & d_{ij}<r.
  \end{cases}
\end{equation*}

\subsection{Numerical Results}
\label{sec:results}
\begin{figure}
\centering
\includegraphics[width=0.85\columnwidth]{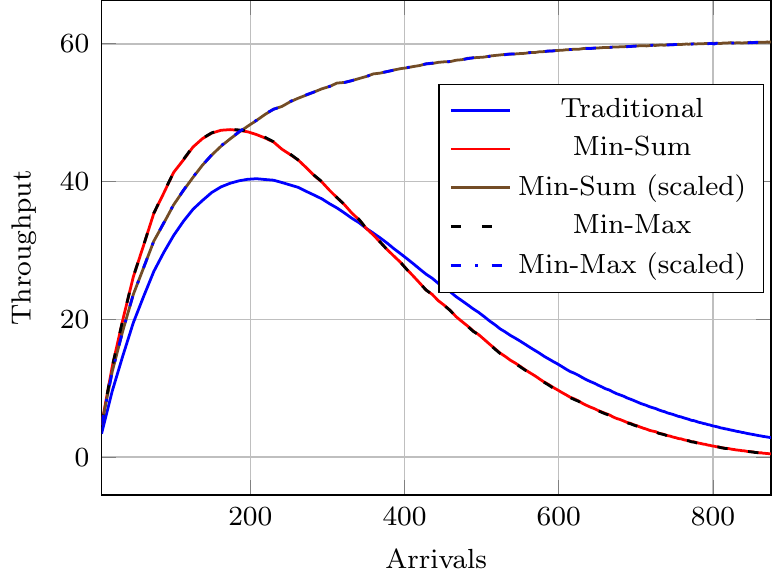}
\caption{Throughput in the spatio-temporal scenario with $150$ slots and varying average arrivals.}
\label{fig:results_point}
\vspace{-6pt}
\end{figure}

We evaluate the algorithms in a system with a fixed number of $N=1000$ users, and compare the throughput to the traditional random access scenario where users transmit in a slot drawn from a uniform distribution.
\Cref{fig:results_point} shows the throughput within a square area with side lengths $L=100$, correlation radius $r=15$ and $K=150$ slots. The global event rate $\lambda$ is varied from $\lambda=0.1/(L^2)$ to $\lambda=30/(L^2)$. The schemes perform equivalently when the number of arrivals is small since collisions are unlikely. However, as the number of arrivals increases, the correlation based schemes perform significantly better than the traditional case, and achieves maximum throughput when the average arrivals is close to the number of slots. When the number of arrivals increases beyond the number of slots, collisions are unavoidable and the throughput decreases significantly. However, with the scaling heuristic only a (random) subset of the users will transmit, and the high throughput is maintained. Although the actual performance of the scaling heuristic depends on the user correlation, the results suggest that it works as intended when the system is under high load.

\section{Practical Aspects and Future Work}
\label{sec:practicalaspects}
Throughout the paper we have assumed that all product expectations are known, and we have ignored the aspect of control overhead involved in scheduling the users. In most practical systems, the expectations need to be learned and it may be desired to limit the control overhead.
The expectations can be learned using maximum likelihood estimation or Bayesian methods, although these may be challenged by the curse of dimensionality imposed by the high number of users. For a high number of users, it may be more suitable to apply methods from data mining, such as frequent itemset mining, where items that frequently occur together are tracked in an online manner~\cite{giannella2003mining}. Since frequent itemset mining only keeps track of the users that transmit most frequently together, it will not provide correlation estimates for all pairs of users. However, as the pairwise correlations are likely to be very sparse, this provides a way of compressing the estimates, assuming the remaining infrequent correlations to be zero.

The proposed random access scheme induces a certain overhead 
to schedule the users. If the users are assigned single slots, it requires at least $N\log_2(K)$ bits assuming that the exact number of users and slots are known to the users, while in the general case where the users are assigned slot transmission probabilities at least $N(K-1)P$ bits where $P$ is the number of bits used to encode a slot transmission probability and $K-1$ reflects the degrees of freedom in the slot assignment. However, in practical systems where the correlation is sparse, it is possible to reduce the amount of signaling by only scheduling some of the users. Suppose as an example that $M\ll N$ users are scheduled, then only $M\log_2(NK)$ bits are needed in the single slot case and $M(K-1)P\log_2(N)$ bits in the probabilistic case.
Furthermore, if the activity changes over time, the information needs to be signaled more often. In this case, the users need to be rescheduled regularly. However, estimating new correlations when the users are already scheduled is challenging since the system faces the classical trade-off between exploration and exploitation. To this end, one may use reinforcement learning and the multi-armed bandit framework.

\section{Conclusion}
\label{sec:conclusion}
This paper has studied how correlated users affect random access protocols, and how information about the correlation can be exploited in the design of random access protocols. We present two algorithms that attempt to maximize the throughput using pairwise correlation information, and evaluate them in two scenarios with correlated user activity. We show that the presented algorithms achieve considerably high throughput compared to traditional random access schemes. This suggests that taking correlation information into account in the random access protocols is promising in scenarios where users are likely to be correlated, such as in MTC.

\section*{Acknowledgment}
This work has been supported in part by WILLOW (ERC Consolidator Grant no. 648382) and TACTILENet (Grant no. 690893), within the Horizon 2020 Program.

\bibliographystyle{IEEEtran}

\end{document}

%% file: motivating_example_lambda1.tex
\begin{tikzpicture}
\begin{axis}[no markers,grid=both,
    title style={yshift=-7pt},
font=\tiny,
xlabel={$p_{11}$},
ylabel={TP},
title={$\lambda=1.0$},
xmax=0.7,
width=4.9cm,
height=5.0cm,
xlabel style={name=xlabel},
legend columns=-1,
legend entries={Traditional, Correlation based},
legend style={legend cell align=left,font=\scriptsize},
legend to name=named
]
\addplot coordinates { (0,1.0) (0.5,0.0) }; 
\addplot coordinates { (0,1) (0.25,0.5) (0.5, 0.5) }; 
\fill[fill=gray!50,opacity=.7] (axis cs:0.25,0.5) -- (axis cs:0.5,0.0) -- (axis cs:0.5,0.5) -- cycle;
\node[label={0:{$(0,\lambda)$}},circle,fill,inner sep=1pt] at (axis cs:0,1) {};
\node[label={180:{$(\lambda/4,\lambda/2)$}},circle,fill,inner sep=1pt] at (axis cs:0.25,0.5) {};
\node[label={90:{$(\lambda/2,\lambda/2)$}},circle,fill,inner sep=1pt] at (axis cs:0.5,0.5) {};
\node[label={180:{$(\lambda/2,0)$}},circle,fill,inner sep=1pt] at (axis cs:0.5,0) {};

\end{axis}
\end{tikzpicture}

%% file: motivating_example_lambda11.tex
\begin{tikzpicture}
\begin{axis}[no markers,grid=both,
    title style={yshift=-7pt},
font=\tiny,
xlabel={$p_{11}$},
ylabel={TP},
title={$\lambda=1.1$},
xmax=0.7,
ymax=1,
width=4.9cm,
height=5.0cm
]
\addplot coordinates { (0.1,0.9) (0.55,0.0) }; 
\addplot coordinates { (0.1,0.9) (0.275,0.55) (0.55,0.55) }; 
\fill[fill=gray!50,opacity=.7] (axis cs:0.275,0.55) -- (axis cs:0.55,0.0) -- (axis cs:0.55,0.55) -- cycle;
\node[label={0:{$(\lambda-1,2-\lambda)$}},circle,fill,inner sep=1pt] at (axis cs:0.1,0.9) {};
\node[label={90:{$(\lambda/4,\lambda/2)$}},circle,fill,inner sep=1pt] at (axis cs:0.275,0.55) {};
\node[label={90:{$(\lambda/2,\lambda/2)$}},circle,fill,inner sep=1pt] at (axis cs:0.55,0.55) {};
\node[label={180:{$(\lambda/2,0)$}},circle,fill,inner sep=1pt] at (axis cs:0.55,0) {};
\end{axis}
\end{tikzpicture}

%% file: venndiagram.tex
\def\firstcircle{(0,1.3cm) circle (1.3cm)}
\def\secondcircle{(0,0) circle (1.3cm)}
\def\thirdcircle{(1.3cm,0cm) circle (1.3cm)}
\def\fourthcircle{(1.3cm,1.3cm) circle (1.3cm)}
\begin{tikzpicture}[fill=gray]
  \tikzstyle{every node}=[font=\footnotesize]
  \draw[pattern=north west lines] \firstcircle;
  \fill[fill=white] \secondcircle;
  \fill[fill=white] \thirdcircle;
  \fill[fill=white] \fourthcircle;
  \begin{scope}[opacity=0.5]
    \begin{scope}
      \clip \firstcircle;
      \fill \secondcircle;
      \fill \thirdcircle;
      \fill \fourthcircle;
    \end{scope}
    \draw \firstcircle;
    \draw \secondcircle;
    \draw \thirdcircle;
    \draw \fourthcircle;
  \end{scope}
  \node[above left] at (0,1.3cm) {$E[x_1]$};
  \node[below left] at (0,0) {$E[x_2]$};
  \node[below right] at (1.3cm,0) {$E[x_3]$};
  \node[above right] at (1.3cm,1.3cm) {$E[x_4]$};
  \node[left] at (0.2cm,0.65cm) {$E[x_1x_2]$};
  \node[rotate=45] at (0.65cm,0.65cm) {$E[x_1x_3]$};
  \node[above] at (0.65cm,1.3cm) {$E[x_1x_4]$};
\end{tikzpicture}

%% file: illustrative_algo.tex
\begin{tikzpicture}
  \tikzstyle{every node}=[font=\footnotesize]
  \begin{scope}
    \begin{scope}[yshift=2.2cm,xshift=0.5cm]
    \draw[fill=black!40] (0,0) rectangle +(0.2,0.2);
    \draw[fill=black!40] (0.2,0) rectangle +(0.2,0.2);
    \draw[fill=black!40] (0.4,0) rectangle +(0.2,0.2);
    \draw[fill=black!40] (0.6,0) rectangle +(0.2,0.2);
    \draw[fill=black!40] (0,0.3) rectangle +(0.2,0.2);
    \draw (0.2,0.3) rectangle +(0.2,0.2);
    \draw (0.4,0.3) rectangle +(0.2,0.2);
    \draw[fill=black!40] (0.6,0.3) rectangle +(0.2,0.2);
    \draw[fill=black!40] (0,0.6) rectangle +(0.2,0.2);
    \draw (0.2,0.6) rectangle +(0.2,0.2);
    \draw[fill=black!40] (0.4,0.6) rectangle +(0.2,0.2);
    \draw (0.6,0.6) rectangle +(0.2,0.2);
    \draw[fill=black!40] (0,0.9) rectangle +(0.2,0.2);
    \draw[fill=black!40] (0.2,0.9) rectangle +(0.2,0.2);
    \draw (0.4,0.9) rectangle +(0.2,0.2);
    \draw (0.6,0.9) rectangle +(0.2,0.2);
    \node[anchor=east] at (0,0.1) {user 4};
    \node[anchor=east] at (0,0.4) {user 3};
    \node[anchor=east] at (0,0.7) {user 2};
    \node[anchor=east] at (0,1.0) {user 1};
    \node[anchor=west] at (0.8,0.55) {$\ldots$};
    \draw[->] (0,-0.1) -- (0.9,-0.1) node[right] {time};
    \end{scope}

    \node[draw, circle] (c1) at (0,1.5) {$1$};
    \node[draw, circle] (c2) at (1.5,1.5) {$2$};
    \node[draw, circle] (c3) at (0,0) {$3$};
    \node[draw, circle] (c4) at (1.5,0) {$4$};
    \draw (c1) -- (c2) node[midway,above] {$1/4$};
    \draw (c1) -- (c3) node[midway,left] {$1/4$};
    \draw (c1) -- (c4) node[pos=0.1,right] {$1/2$};
    \draw (c2) -- (c3) node[pos=0.4,right=-2] {$1/4$};
    \draw (c2) -- (c4) node[midway,right] {$1/2$};
    \draw (c3) -- (c4) node[midway,below] {$1/2$};

    \begin{scope}[xshift=-0.0cm,yshift=-3.0cm]
      \node[draw, circle] (s1c1) at (-0.4,1.2) {$1$};
      \node[draw, circle] (s1c2) at (0.0,1.2) {$2$};
      \node[draw, circle] (s1c3) at (0.4,1.2) {$3$};
      \node[draw, circle] (s1c4) at (0,0) {$4$};
      \draw (s1c2) -- (s1c4);
      \node[anchor=north] at (0,-0.2) {$\mathrm{TP}=7/4$};
      \node[anchor=east] at (-0.7,1.2) {Slot 1};
      \node[anchor=east] at (-0.7,0) {Slot 2};
    \end{scope}
    \begin{scope}[xshift=1.5cm,yshift=-3.0cm]
      \node[draw, circle] (s2c1) at (-0.2,1.2) {$1$};
      \node[draw, circle] (s2c2) at (0.2,1.2) {$2$};
      \node[draw, circle] (s2c3) at (-0.2,0) {$3$};
      \node[draw, circle] (s2c4) at (0.2,0) {$4$};
      \draw (0.0,1.0) -- (0.0,0.2);
      \node[anchor=north] at (0,-0.2) {$\mathrm{TP}=1$};
    \end{scope}
    \draw[->] (0.6,-0.4) -- (0,-1.4) node[midway,left] {Min-Max};
    \draw[->] (0.9,-0.4) -- (1.5,-1.4) node[midway,right] {Min-Sum};
  \end{scope}

  \begin{scope}[xshift=4cm]
    \begin{scope}[yshift=2.2cm,xshift=0.3cm]
      \draw[fill=black!40] (0,0) rectangle +(0.2,0.2);
      \draw[fill=black!40] (0.2,0) rectangle +(0.2,0.2);
      \draw[fill=black!40] (0.4,0) rectangle +(0.2,0.2);
      \draw[fill=white] (0.6,0) rectangle +(0.2,0.2);
      \draw[fill=black!40] (0.8,0) rectangle +(0.2,0.2);
      \draw[fill=black!40] (1.0,0) rectangle +(0.2,0.2);
      \draw[fill=black!40] (0,0.3) rectangle +(0.2,0.2);
      \draw[fill=black!40] (0.2,0.3) rectangle +(0.2,0.2);
      \draw[fill=black!40] (0.4,0.3) rectangle +(0.2,0.2);
      \draw[fill=black!40] (0.6,0.3) rectangle +(0.2,0.2);
      \draw[fill=white] (0.8,0.3) rectangle +(0.2,0.2);
      \draw[fill=white] (1.0,0.3) rectangle +(0.2,0.2);
      \draw[fill=white] (0,0.6) rectangle +(0.2,0.2);
      \draw[fill=black!40] (0.2,0.6) rectangle +(0.2,0.2);
      \draw[fill=white] (0.4,0.6) rectangle +(0.2,0.2);
      \draw[fill=black!40] (0.6,0.6) rectangle +(0.2,0.2);
      \draw[fill=white] (0.8,0.6) rectangle +(0.2,0.2);
      \draw[fill=black!40] (1.0,0.6) rectangle +(0.2,0.2);
      \draw[fill=black!40] (0,0.9) rectangle +(0.2,0.2);
      \draw[fill=white] (0.2,0.9) rectangle +(0.2,0.2);
      \draw[fill=black!40] (0.4,0.9) rectangle +(0.2,0.2);
      \draw[fill=white] (0.6,0.9) rectangle +(0.2,0.2);
      \draw[fill=black!40] (0.8,0.9) rectangle +(0.2,0.2);
      \draw[fill=white] (1.0,0.9) rectangle +(0.2,0.2);
      \node[anchor=east] at (0,0.1) {user 4};
      \node[anchor=east] at (0,0.4) {user 3};
      \node[anchor=east] at (0,0.7) {user 2};
      \node[anchor=east] at (0,1.0) {user 1};
      \node[anchor=west] at (1.2,0.55) {$\ldots$};
      \draw[->] (0,-0.1) -- (1.3,-0.1) node[right] {time};
    \end{scope}

    \node[draw, circle] (c1) at (0,1.5) {$1$};
    \node[draw, circle] (c2) at (1.5,1.5) {$2$};
    \node[draw, circle] (c3) at (0,0) {$3$};
    \node[draw, circle] (c4) at (1.5,0) {$4$};
    \draw (c1) -- (c2) node[midway,above] {$0$};
    \draw (c1) -- (c3) node[midway,left] {$1/3$};
    \draw (c1) -- (c4) node[pos=0.1,right] {$0.5$};
    \draw (c2) -- (c3) node[pos=0.4,right=-2] {$1/3$};
    \draw (c2) -- (c4) node[midway,right] {$1/3$};
    \draw (c3) -- (c4) node[midway,below] {$0.5$};

    \begin{scope}[xshift=-0.0cm,yshift=-3.0cm]
      \node[draw, circle] (sc1) at (-0.4,1.2) {$1$};
      \node[draw, circle] (sc2) at (0.0,1.2) {$2$};
      \node[draw, circle] (sc3) at (0.4,1.2) {$3$};
      \node[draw, circle] (sc4) at (0,0) {$4$};
      \draw (sc2) -- (sc4);
      \node[anchor=north] at (0,-0.2) {$\mathrm{TP}=7/6$};
    \end{scope}
    \begin{scope}[xshift=1.5cm,yshift=-3.0cm]
      \node[draw, circle] (sc1) at (-0.2,1.2) {$1$};
      \node[draw, circle] (sc2) at (0.2,1.2) {$2$};
      \node[draw, circle] (sc3) at (-0.2,0) {$3$};
      \node[draw, circle] (sc4) at (0.2,0) {$4$};
      \draw (0.0,1.0) -- (0.0,0.2);
      \node[anchor=north] at (0,-0.2) {$\mathrm{TP}=9/6$};
    \end{scope}
    \draw[->] (0.6,-0.4) -- (0,-1.4) node[midway,left] {Min-Max};
    \draw[->] (0.9,-0.4) -- (1.5,-1.4) node[midway,right] {Min-Sum};
  \end{scope}
\end{tikzpicture}

%% file: spatiotemp_traffic_model.tex
\begin{tikzpicture}
\begin{axis}[
    domain=0:4,
    xtick=\empty, ytick=\empty,
    xmin=0, xmax=4,
    ymin=0, ymax=4,
    width=6cm,
    height=6cm
]
\draw[dashed] (axis cs:1.5,2.5) circle [blue, radius=1];
\draw[->] (axis cs:1.5,2.5) -- (axis cs:2.5,2.5) node[midway,above] {$r$};
\node[circle,fill=black,inner sep=1.5pt] at (axis cs:1.5,2.5) {};
\addplot [red, only marks, mark=*, samples=20, mark size=0.75, variable=\t, domain=0:2*pi]
    ( {cos(deg(t))+1.5}, {2.5+sin(deg(t))-2*rnd*(sin(deg(t)))} );

\addplot [blue, only marks, mark=*, samples=10, mark size=0.75, variable=\t, domain=0:pi]
    ( {cos(deg(t))+1.5}, {2.5+sin(deg(t))+rnd*(2-0.5-sin(deg(t)))} );
\addplot [blue, only marks, mark=*, samples=40, mark size=0.75, variable=\t, domain=0:pi]
    ( {cos(deg(t))+1.5}, {2.5-sin(deg(t))-rnd*(2+0.5-sin(deg(t)))} );
\addplot [blue, only marks, mark=*, samples=10, mark size=0.75, domain=0:0.5]
    {2+rand*2};
\addplot [blue, only marks, mark=*, samples=60, mark size=0.75, domain=2.5:4]
    {2+rand*2};
  \end{axis}
  \draw[|-|] (0,-0.2) -- (4.42cm,-0.2) node[midway,below] {$L$};
  \draw[|-|] (-0.2,0) -- (-0.2,4.42cm) node[midway,left] {$L$};
\end{tikzpicture}